\documentclass{aa}
\usepackage{graphicx}
\usepackage{amssymb}
\usepackage{latexsym}
\usepackage{epsfig}
\usepackage{txfonts}

\def\rs{\rm s}
\def\rs1{\rm s^{-1}}

\def\rcm{\rm cm}
\def\rcm2{\rm cm^{-2}}

\def\c2r{\chi^2_\nu}

\def\epo{$E_{\rm p}$ }
\def\epi{\ensuremath{E_{\rm p,i}}}
\def\eiso{\ensuremath{E_{\rm iso}}}
\def\ega{\ensuremath{E_{\rm jet}}}
\def\epeiso{$E_{\rm p,i}$ -- $E_{\rm iso}$}

\def\epega{$E_{\rm p,i}$ -- $E_{\gamma}$ }

\def\apjs{{\it Astrophys. J. Suppl. Ser. }}

\begin{document}

\title{On the consistency of peculiar GRBs 060218 and 060614
with the \epeiso{} correlation}

\author{L.~Amati\inst{1}
\and M.~Della Valle\inst{2,3}
\and F.~Frontera\inst{3,1}
\and D.~Malesani\inst{4}
\and C.~Guidorzi\inst{5,6}
\and E.~Montanari\inst{3,7}
\and E.~Pian\inst{8}
}

\offprints{L. Amati: \email{amati@iasfbo.inaf.it}}

\institute{INAF - Istituto di Astrofisica Spaziale e Fisica Cosmica Bologna,
via P. Gobetti 101, I-40129 Bologna, Italy 
\and
INAF - Osservatorio Astrofisico di Arcetri, largo E. Fermi 5, I-50125 Firenze, Italy
\and
Dipartimento di Fisica, Universit\`a di Ferrara, Via Paradiso 12, I-44100 Ferrara, Italy
\and
International School for Advanced Studies (SISSA-ISAS), via Beirut 2-4, I-34014 Trieste, Italy
\and
Dipartimento di Fisica, Universit\`a degli Studi di Milano Bicocca, piazza delle Scienze 3, I-20126 Milano, Italy
\and
INAF - Osservatorio Astronomico di Brera, via E. Bianchi 46, I-23807 Merate (LC), Italy
\and
ITA ``Calvi'', via Digione 20, I-41034 Finale Emilia (MO), Italy
\and
INAF - Osservatorio Astronomico di Trieste, via G. Tiepolo 11, I-34131 Trieste, Italy
}

\date{Received; Accepted }
 
\abstract{We analyze and discuss the position of GRB\, 060218 and
GRB\,060614 in the \epeiso{} plane. GRB\,060218 is important because
of its similarity with GRB 980425, the proto--type event of the
GRB--SN connection. While GRB\,980425 is an outlier of the \epeiso{}
correlation, we find that GRB\,060218 is fully consistent with
it. This evidence, combined with the `chromatic' behavior of the
afterglow light curves, is at odds with the hypothesis that GRB 060218
was a `standard' GRB seen off-axis and supports the existence of a
class of truly sub--energetic GRBs.\\ GRB\,060614 is a peculiar event
not accompanied by a bright Supernova. Based on published spectral
information, we find that also this event is consistent with the
\epeiso{} correlation. We discuss the implications of our results for
the rate of sub--energetic GRBs, the GRB/SN connection and the
properties of the newly discovered sub--class of long GRBs not
associated with bright Supernovae. We have included in our analysis
other recent GRBs with clear evidence (or clear no evidence) of
associated SNe.
}

\authorrunning{Amati et al.}
\titlerunning{Consistency of GRB\,060218 and GRB\,060614 
with the \epeiso{} correlation}

\maketitle

\begin{figure*}
\centerline{\epsfig{figure=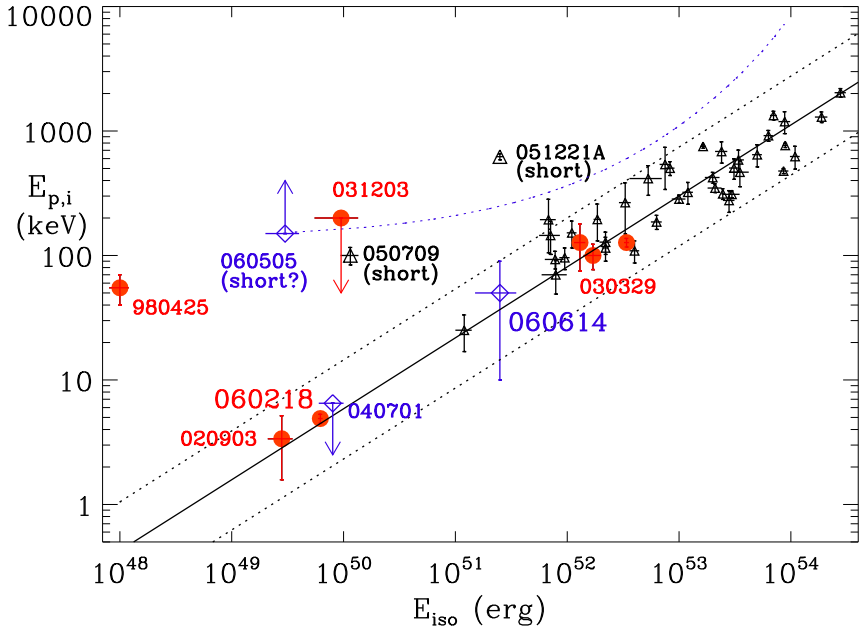,width=13cm,angle=0}}
\caption{Distribution of \epi{} and \eiso{} values of GRBs and XRFs
with firm estimates of $z$ and \epi{}, including also two short GRBs
with known redshift: GRB\,050709 and GRB\,051221A. Data and models are taken from
Amati (2006), except for GRB\,060218, GRB\,040701 and GRB\,060614 (see text).
The solid line shows the best--fit power--law obtained by
Amati (2006) without
including GRB\,060218, GRB\,980425 and GRB\,031203; the two parallel dotted lines
indicate the 2-$\sigma$ confidence region. 
Those GRBs with evidence of association with a SN (Table 1) are marked with
big dots. The location in the \epeiso{} plane of GRB\,060614 and other the two 
events with deep limits
to the magnitude of the associated SN, XRF\,040701 and GRB\,060505, are 
shown as big diamonds. The curved dotted line shows how the GRB\,060505 
point moves in the
\epeiso{} plane as a function of redshift.
}
\end{figure*}

\section{Introduction}

Almost a decade of optical, infrared and radio observations of
gamma-ray bursts (GRBs) has allowed to link long-duration GRBs
(or, at least, a fraction of them) with the death of massive
stars. This result is based on three pieces of evidence: i) there are
(to date) four clear cases of association between ``broad lined''
supernovae (BL-SNe) (i.e. SNe-Ib/c characterized by a large kinetic
energy, often labeled as hypernovae, HNe hereafter) and GRBs:
GRB\,980425/SN\,1998bw (Galama et al. 1998), GRB\,030329/SN\,2003dh
(\cite{Stanek03,Hjorth03}), GRB\,031203/SN\,2003lw
(Malesani et al. 2004) and GRB\,060218/SN\,2006aj (Masetti et
al. 2006, Campana et al. 2006; Pian et al. 2006); ii) in a few cases,
spectroscopic observations of bumps observed during the late decline
of GRB afterglows has revealed the presence of SN features (\cite{
Dellavalle03,Dellavalle06a,Soderberg05}); iii) long GRBs are located inside star
forming galaxies (\cite{Djorgovski98,LeFloch03,
Christensen04,Fruchter06}). The standard
theoretical scenario suggests
that long GRBs are produced in the collapse of the core of H/He
stripped-off massive stars (possibly Wolf--Rayet, see Campana et
al. 2006) with an initial mass higher than $\sim 20$ M$_{\sun}$ and
characterized by a very high rotation speed (e.g. \cite{Woosley93,Paczynski98}).\\
GRB980425 was not only the first example of the GRB--SN
connection, but also a very peculiar event. Indeed, with a redshift of
0.0085 it was much closer than the majority of GRBs with known redshift
($\sim0.1<z<6.3$) and its total radiated energy under the assumption of
isotropic emission, \eiso, was very low ($\sim$10$^{48}$ erg),
therefore well below the typical range for ``standard'' bursts
($\sim$10$^{51}$ -- $\sim$10$^{54}$ erg).  Moreover, this event was
characterized by values of \epi, the photon energy at which the
$\nu$F$_\nu$ spectrum peaks (hence called {\it peak energy}), and \eiso{}
completely inconsistent with the \epi $\propto$ \eiso$^{0.5}$
correlation holding for long ``cosmological'' GRBs (Amati et
al. 2002).\\
This correlation has not only several implications for the physics,
jet structure and GRBs/XRFs unification scenarios, but
can be used to investigate the
existence of different sub--classes of GRBs
(e.g. \cite{Amati06}). 
In addition to GRB\,980425, also GRB\,031203/SN\,2003lw
(\cite{Sazonov04,Malesani04}) was characterized by a value of \epi{} which,
combined with its low value of \eiso{}, makes it the second (possible)
outlier of the \epeiso{} correlation (the \epi{} value of this event is
still debated). Both cases may point towards the existence of a class of
nearby and intrinsically faint GRBs with different properties with
respect to ``standard'' GRBs. However, it has been suggested by several authors
that the low measured \eiso{} of these events and their inconsistency with the 
\epeiso{} correlation are due to viewing angle effects (off--axis
scenarios, see, e.g., \cite{Yamazaki03,Ramirez05}). 

In this paper we focus on the the position, in the \epeiso{} plane, of
two recently discovered events: GRB\,060218 and GRB\,060614.
GRB\,060218 is particularly important because of its association with
SN\,2006aj at $z = 0.033$
(\cite{Masetti06,Campana06,Soderberg06a,Pian06,Modjaz06,Sollerman06,Mirabal06,Cobb06,Ferrero06}). In
addition, this GRB event was both ``local'' and ``sub-energetic'' like
GRB\,980425 and GRB\,031203, but unlike them it matches the \epi{}
vs. \eiso{} relationship.  GRB\,060614 is very interesting because
of the very deep upper limits to the luminosity of the possible
associated SN (\cite{Dellavalle06b, Fynbo06,Gal-Yam06}).  We find
that also this event is consistent with the \epeiso{} correlation
(section 3.4). Our analysis includes also other GRBs with evidence for
associated SNe and two nearby GRBs which are not accompanied by bright
SN explosions. We discuss the implications of our results for the
existence and rate of sub--energetic GRBs, the GRB/SN connection and
the properties of the sub--class of long GRBs not associated with
bright Supernovae.\\ Throughout this paper we assumed H$_0$=70 km
s$^{-1}$ Mpc$^{-1}$, $\Omega_M$ = 0.3 and $\Omega_{\Lambda}$ = 0.7.

\begin{center}
\begin{table*}[t!]
\caption {Upper panel: properties of GRBs with known $z$ and associated SN; the
first four bursts are those most clearly associated with a SN event,
the following three are those GRBs with firm estimates of \epi{} and
evidence of SN features in the spectrum of the late-time optical
afterglow. Lower panel: events with deep limits to the magnitude of a
possible associated SN.  The values of \epi{} and \eiso{} are taken
from Amati (2006), except for GRB\,060218, GRB\,040701, GRB\,060505
and GRB\,060614 (see text); those of $\theta_{\rm jet}$ and \ega{} are
taken from Nava et al. (2006) and Friedman \& Bloom (2005), except for
GRB\,060218 (\cite{Soderberg06b,Fan06}) and GRB\,060614 (Della Valle
et al. 2006).  The \epi{} upper limit for GRB\,031203 is based on the
value by Ulanov et al. (2005) combined with the evidence of a soft
X--ray excess inferred from the dust echo measured by XMM (see text).
References for SN properties are given in the last column. M$_V$
of SN 2005nc was derived after assuming (B--V)$\sim$0.5; 
M$_V$ of SN 2002lt was derived after assuming (U--V)$_{max}\sim 0.2$, 
as observed for SN 1994I (\cite{Richmond96}).}
\begin{tabular}{lllllllll}
\hline
\hline 
GRB/SN & $z$ & \epi & $E_{\rm prompt}^{\rm iso}$ & $\theta_{\rm jet}$ &$E_{\rm prompt}^{\rm jet}$  & SN $E_{\rm K}^{\rm iso}$$^{(a)}$ &
SN peak mag & Ref.$^{(b)}$ \\
 &  &  (keV)  & (10$^{50}$ erg) & (deg) & (10$^{50}$ erg) & (10$^{50}$ erg) &  &  \\
\hline
980425/SN\,1998bw & 0.0085& 55$\pm$21 & 0.01$\pm$0.002 & - & $<$0.012 & 200-500 &$M_V$=$-$19.2$\pm$0.1 & (1,2,3,4)  \\
060218/SN\,2006aj & 0.033 & 4.9$\pm$0.3 & 0.62$\pm$0.03 & $>$57 & 0.05--0.65 & 20--40 & $M_V$=$-$18.8$\pm$0.1  & (5,6)\\
031203/SN\,2003lw & 0.105 & $<$200  & 1.0$\pm$0.4 & -- & $<$1.4 & 500-700 & $M_V$=$-$19.5$\pm$0.3  & (3,7)\\
030329/SN\,2003dh & 0.17  & 100$\pm$23 & 170$\pm$30 & 5.7$\pm$0.5 & 0.80$\pm$0.16 & $\sim$400 & $M_V$=$-19.1\pm 0.2$  & (3,8)\\
020903/BL-SNIb/c  & 0.25  & 3.4$\pm$1.8 & 0.28$\pm$0.07 & -- & $<$0.35 & --  & $M_V\sim-18.9$ & (9) \\
050525A/SN\,2005nc& 0.606 & 127$\pm$10 & 339$\pm$17  & 4.0$\pm$0.8 & 0.57$\pm$0.23 & --. & 
$M_V$=$-19.4^{+0.1}_{-0.5}$ & (10)\\
021211/SN\,2002lt & 1.01 & 127$\pm$52 & 130$\pm$15 & 8.8$\pm$1.3 & 1.07$\pm$0.13 & --  & 
$M_V\sim-19$$\pm$1 & (11)\\
\hline
060505 & 0.089 & $>$160 & 0.3$\pm$0.1  & -- & -- & -- & $M_V > -13.5$ & (12)\\
060614 & 0.125 & 10--100 & 25$\pm$10 & $\sim$12 & 0.45$\pm$0.20 & --  & $M_V > \sim-13$ & (12,13,14)\\
040701 & 0.215 & $<$6. & 0.8$\pm$0.2  & -- & -- & -- & $M_V > -16$ & (9)\\

\hline
\hline
\end{tabular}
\begin{list}{}{}
\item[$^{(a)}$] Values derived by modeling optical data of the SN component with
hypernova models, like, e.g., the 1--dimensional synthesis code by \cite{Mazzali06a}. 
\item[$^{(b)}$] References: (1) \cite{Mazzali01}, (2) \cite{Maeda06}, (3) \cite{Mazzali06a}
and references therein,
(4) \cite{Galama98}, (5) \cite{Mazzali06b}, (6) \cite{Pian06}, (7) \cite{Malesani04},
(8) \cite{Hjorth03}, (9) \cite{Soderberg05}, (10) \cite{Dellavalle06a}, (11)
\cite{Dellavalle03}, (12) Gal--Yam et al. (2006), (13) Della Valle et al. (2006b),
(14) Fynbo et al. (2006).
\end{list} 
\end{table*}
\end{center}

\section{Peculiar GRBs in the \epeiso{} plane}

\subsection{GRB\,060218, sub--energetic GRBs and GRB/SN events}

GRB\,060218 was detected by Swift/BAT on 2006 February 18, at 03:34:30
UT and fast pointed and localized by Swift/XRT and UVOT
(\cite{Cusumano06a}). The prompt event was anomalously long ($T_{90}$
= 2100$\pm$100 s) and very soft (average 15--150 keV photon index of
$\sim$2.5), with a 15--150 keV fluence of $(6.8 \pm 0.4) \times
10^{-6}$ erg cm$^{-2}$ (\cite{Sakamoto06}). The spectrum of the host
galaxy showed narrow emission lines at a redshift $z = 0.033$
(\cite{Mirabal06}), whereas the optical counterpart showed a blue
continuum and broad spectral features characteristic of a supernova
(\cite{Pian06,Modjaz06,Sollerman06,Ferrero06}). Similarly to
GRB\,980425 and GRB\,031203, GRB\,060218 exhibited a very low
afterglow kinetic energy ($\sim 100$ times less than standard GRBs),
as inferred from radio observations (\cite{Soderberg06b}). Based on
Swift/BAT and XRT preliminary results (\cite{Sakamoto06,Cusumano06b}),
Amati et al.{} (2006) argued that the GRB\,060218 properties were
consistent with the \epeiso{} correlation.
This result is confirmed after adopting the refined Swift/XRT and BAT
data (\cite{Campana06}). Fig.~1 shows the position of GRB\,060218 in
the \epi{} vs \eiso{} plane to be fully consistent with the \epeiso{}
correlation.  When adding this event to the sample of Amati (2006),
the Spearman rank correlation coefficient between \epi{} and \eiso{}
turns out to be 0.894 (for 42 events), corresponding to a chance
probability as low as $\sim 2 \times 10^{-15}$.  Fig.~1 also shows
that another very soft and weak event, XRF\,020903, matches the
\epeiso{} correlation (\cite{Sakamoto04}). Thus, XRF\,020903 and
GRB\,060218 may simply represent the extension to low energy ($\eiso <
10^{51}$ erg) of the ``cosmological'' GRB sequence.  In addition,
based on the lack of a break in the radio light curve, a lower limit
of 1--1.4 rad can be set to $\theta_{\rm jet}$ (Soderberg et
al. 2006).  This value is much higher than those of classical,
cosmological GRBs (e.g. Nava et al. 2006), further supporting the idea
that close-by, sub--energetic GRBs have a much less collimated emission
(\cite{Soderberg06b,Guetta06}). This also
implies that the collimation--corrected energy, \ega{}, released
during prompt emission is not much lower than \eiso{}, lying in the
range $(\sim2.7\mbox{--}6.5) \times 10^{49}$ erg.\\ A different (well
known) behaviour is exhibited by GRB\,980425. Less straightforward is
the interpretation of the position of GRB\,031203.  Based on the
detection by XMM--Newton of a transient dust--scattered X-ray halo
associated with it, some authors
(\cite{Vaughan04,Ghisellini06,Watson06,Tiengo06}) argued that this
event might have been much softer than inferred from INTEGRAL/ISGRI
data (\cite{Sazonov04}). Finally, we plot in Fig. 1 
also short GRBs with known redshift (namely
GRB\,050709 and GRB\,051221A) which lie outside of the region
populated by long events (see also Amati 2006).

\subsection{GRB\,060614 and other no--hypernova events}

GRB\,060614 was detected by Swift/BAT on June 14, 2006 at 12:43:48 UT
as a long (120 s) event showing a bright initial flare followed by
softer, extended prompt emission (\cite{Parsons06}).  Follow--up
observations of the bright X--ray and optical counterparts detected
and localized by XRT and UVOT led to the identification of an host
galaxy lying at $z$ = 0.125 (\cite{Price06,Fugazza06}).  GRB\,060614
is a very important event, because the upper limit to the luminosity
of the SN possibly associated with it was at least two orders of
magnitude fainter (\cite{Dellavalle06b,Fynbo06,Gal-Yam06}) than the
peak luminosity of broad-lined SNe-Ibc normally associated with GRBs
(see Table 1).  In order to analyze the location of GRB\,060614 in the
\epeiso{} plane, we performed estimates of \epi{} and \eiso{} based on
the spectra and fluences of two portions of the event measured by
Konus--Wind in 20 keV -- 2 MeV energy range.  Golenetskii et
al. (2006) report \epo of 302$_{-85}^{+214}$ keV for the first bright
pulse lasting $\sim$8.5 s and providing $\sim$ 20\% of the total
fluence.  The Konus--Wind spectrum of the subsequent part of the
event, composed by softer pulses, can be fitted by a simple power--law
with photon index of $\sim$2.13$\pm$0.05, indicating that \epo{} maybe
close or below the 20 keV low energy bound. In order to estimate a
reasonable \epi{} range for the average spectrum, we performed both a
weighted average of the \epo{} measured in the two time intervals (by
assuming \epo 10 or 20 keV for the second interval) and simulations
(i.e. we generated fake spectra of the two intervals, summed them and
fit them with the Band function). We find that \epi{} of GRB\,060614
likely lies in the range 10--100 keV, which is also consistent with
the average photon index of $\sim$2 measured by Swift/BAT in 15--150
keV (\cite{Barthelmy06}, Mangano et al. in preparation).  As can be
seen in Fig.~1, GRB\,060614 is consistent with the \epeiso{}
correlation as most GRB/SN events, therefore suggesting that the
position in the \epeiso{} plane of long GRBs does not critically
depend on the progenitors properties. \\ In the second part of Table
1, we also report the \epi{} and \eiso{} values, or upper/lower
limits, for other two long GRBs/XRFs with deep limits to the magnitude
of a possible associated SN, XRF\,040701 (\cite{Soderberg05}) and
GRB\,060505 (\cite{Fynbo06}).  For these two events, no estimates of
\epo{} are available and we could estimate only approximate upper /
lower limits based on the available published information. In the case
of XRF\,040701, \eiso{} and the upper limit to \epi{} were inferred
based on the HETE--2 spectral information reported by Barraud et
al. (2004), who quote an average photon index of 2.4$\pm$0.3 . This
indicates that the peak energy of XRF\,040701 is likely towards, or
below, the low bound of the WXM + FREGATE 2--400 keV energy band. We
conservatively assumed \epo{} $<$ 5 keV, which, by assuming the
redshift of 0.215, translates in an upper limit to \epi{} of $\sim$6
keV. The \eiso{} range was computed by assuming a Band spectral shape
with $\alpha = -1.5$, $\beta = -2.4$ and \epo = 1--5 keV, normalized
to the measured 2--30 keV fluence. Poor spectral information is
available for GRB\,060505; based on Swift/BAT data, Hullinger et
al. (2006) report a photon index of 1.3$\pm$0.3 in the 15--150 keV
energy range. Thus, for this event \epo{} is likely above 150 keV;
\eiso{} was computed by assuming a Band spectral shape with $\alpha =
-1.3$, $\beta = -2.5$ and \epo = 150--1000 keV, normalized to the
measured 15--150 keV fluence. Finally, the VLT afterglow light curve
of GRB\,060614 shows a break which, if interpreted as due to
collimated emission, gives a jet angle of $\sim$12 deg (Della Valle et
al. 2006b) and a collimation corrected radiated energy \ega{} of
4.5$\pm$2.0$\times$10$^{49}$ erg, consistent with the \epega{}
correlation (Ghirlanda et al. 2004). \\

\section{Discussion}

\subsection{GRB\,060218: existence of truly sub--energetic GRBs}

The fact that the two closest, sub--energetic, and SN--associated GRBs, 980425 and
031203, are
outliers of the \epeiso{} correlation stimulated  
several works also in the
framework of GRB/SN unification models. The most common interpretation
is that they were ``standard'' GRBs viewed off-axis
(e.g., \cite{Yamazaki03, Ramirez05}). 
These scenarios explain both the low value of \eiso{} and the
deviation from the \epeiso{} correlation by means of relativistic Doppler and beaming 
effects. 
For instance (e.g., \cite{Yamazaki03}), by assuming a uniform jet and a viewing angle,
$\theta$$_{\rm v}$, larger than the jet opening angle, $\theta$$_{\rm jet}$, 
it is found that \epi{} $\propto$ $\delta$ and \eiso{} $\propto$ $\delta^{1-\alpha}$, 
where $\alpha$ is the average power--law index of the prompt emission photon spectrum in
the hard
X--ray energy band (typically between $-$1 and $-$2) and
$\delta$
is the relativistic Doppler factor 
$\delta = \{\Gamma[1 - \beta \cos(\theta_{v} - \theta_{\rm jet})]\}^{-1}$ 
($\Gamma$ is the bulk Lorentz factor of the plasma and $\beta$
is the
velocity of the outflow in units of speed of light), which decreases as  
$\theta_{v}$ increases. For large off--axis
viewing angles the different dependence of \epi{} and \eiso{} on $\delta$ would cause
significant deviations from the \epeiso{} correlation and a very low {\it observed}
value
of \eiso{}. 
Off--axis scenarios make also predictions on the multi--wavelength 
afterglow light curve.
At the beginning, when the Lorentz factor of
the relativistic shell, $\Gamma$, is very high, the flux detected by the observer is much 
weaker with respect to the case $\theta_{\rm v}$ $<$ $\theta_{\rm jet}$ .
As $\Gamma$ decreases, and thus the beaming angle (which is proportional to 1/$\Gamma$)
increases, the observer measures a slow raise, or a flat light curve in case
the GRB is structured (e.g. \cite{Granot02,Rossi02}).
The light curves show a peak or a smooth break
when 1/$\Gamma$ = $\theta_{\rm v}$, and then
behave in the same way as for an on--axis observer. This peak, or break, is due
to a purely geometrical factor, thus it should be "achromatic", i.e. occur at the
same times at all wavelengths.
While theoretical modeling shows that the nebular spectrum of SN1998bw, associated
with GRB\,980425,
is consistent with an aspherical explosion seen off--axis (\cite{Maeda06}),
radio-observations (\cite{Berger03a})
seem to exclude the detection of relativistic off--axis ejecta, because
of the lack of the detection of the late (a few years at most)
radio re-brightening predicted by off--axis models (see also
\cite{Ramirez05}). However, it was suggested that the
still low radio flux may be still consistent with the off--axis interpretation
if the density of the circum--burst wind is at least 1 order of
magnitude lower than expected (e.g., \cite{Waxman04}).

In light of the above arguments,
the consistency of GRB\,060218 with the \epeiso{} correlation, 
as presented in Fig.~1 and 
discussed in Section~2,
suggests that
this event, the most similar to GRB\,980425 because of its very
low \eiso, very low $z$ and prominent association with a SN (2006aj),
was not an event seen off--axis.
This conclusion is further supported by its multi--wavelength afterglow
properties.
The early-time ($t \la 0.5$~d) light curves of GRB\,060218 exhibited a slow
rise, as observed in the optical/UV, X-ray and soft gamma-ray bands
(\cite{Campana06}). However, the peak time was dependent upon the
frequency, occurring earlier at shorter wavelengths, contrary to the
expectations for an off-axis jet (e.g. \cite{Granot02}).
Another piece of evidence comes from
radio data: the radio afterglow light curves can be fitted with standard
GRB afterglow models (i.e. without the need to involve viewing angle
effects), as shown by Soderberg et al. (2006) and Fan, Piran \& Xu
(2006). \\
Finally, as is
shown in Fig.~2, we find that GRB\,060218 and the other sub--energetic
events GRB\,980425 and GRB\,031203 follow and extend the correlation
between \eiso{} and the X-ray afterglow 2--10 keV luminosity at 10 hr
from the event reported by De Pasquale et al. (2006) and Nousek et
al. (2006). For the events in the sample of Nousek et al. (2006)
the 2--10 keV $L_{\rm X,10}$ values were computed from the 0.3-10 keV values 
by assuming a
typical X-ray afterglow photon index of 2 .
The 1--10000 keV \eiso{} values are
taken from Amati (2006);
for those few events not included in the sample of Amati (2006), we derived the 
1--10000
keV \eiso{} from the 10--500 keV value reported by Nousek et al. (2006) by assuming a 
Band spectral
shape (\cite{Band93}) 
with $\alpha$=$-$1, $\beta$=$-$2.3 and E$_0$=15 keV, for those events with 15--150 keV 
photon
index $>$ 2, or E$_0$=200 keV, for those events with photon index $<$ 2 .
The
X-ray luminosities for GRB\,980425, GRB\,031203 and GRB\,060218 were
derived from the X--ray afterglow light curves reported by Pian
et al. (2000), Watson et al. (2004) and Campana et al. (2006),
respectively. 
Fig.~2 clearly shows that these 3 events are
sub--energetic also from the point of view of their X--ray afterglow
emission. While the correlation based on ``normal'' events was found to
be only marginally significant (\cite{DePasquale05,Nousek06}), here we
show that after including sub--energetic GRBs, it becomes highly
significant (chance probability $\sim 10^{-11}$). This result further
indicates that sub--energetic GRBs may be intrinsically weak and belong
to an extension of the normal cosmological events. Also, qualitatively,
in the off--axis viewing scenarios, one would expect that, due to the
rapidly decreasing Lorentz factor of the fireball, and thus to the
rapidly increasing beaming angle, the prompt emission
should be much more depressed with respect to afterglow emission at
$\sim10$ hours. This would imply that the points corresponding to the
three sub--energetic events should lie above the extrapolation of the
law best fitting on--axis GRBs, which is not the case.

All the above evidences indicate that the local under--luminous
GRB\,060218 was not seen off--axis and 
point towards the existence of a class of truly sub--energetic 
GRBs. 

\begin{figure}
\epsfig{figure=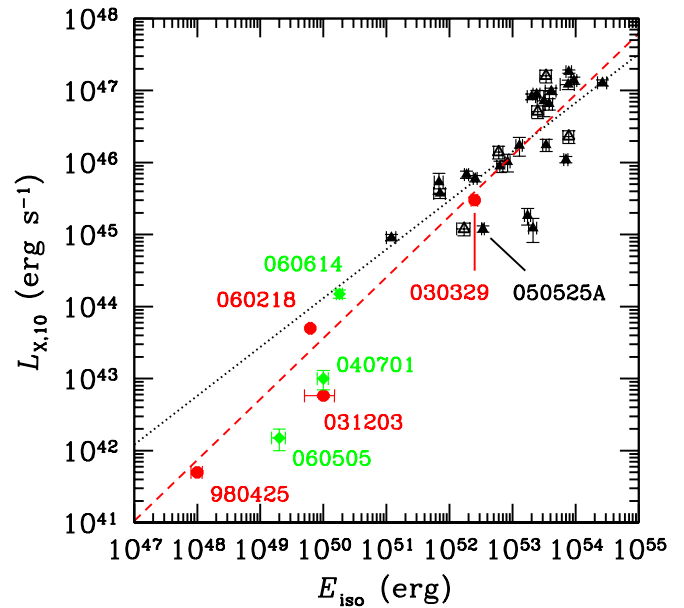,width=\columnwidth}

\caption{2--10 keV afterglow luminosity at 10 hours $L_{\rm X,10}$
vs. \eiso{} for the events included in the sample of Nousek et
al. (2006; triangles) plus the 3 sub--energetic GRB\,980425, GRB\,031203,
GRB\,060218,  the other GRB/SN event GRB\,030329 (circles), and
3 GRBs with known $z$ and deep limits to the peak magnitude
of associated SN, XRF\,040701, GRB\,060505 and GRB\,060614 (diamonds). Empty triangles
indicate those GRBs for which the 1-10000 keV \eiso{} was computed based on 
the 100--500 keV \eiso{} reported by Nousek et al. (2006) by assuming
an average spectral index (see text). 
The plotted lines are the best fit power-laws obtained
without (dotted) and with (dashed) sub--energetic GRBs and GRB\,030329.}
\end{figure}

\subsection{Implications for GRBs occurrence rate}

\begin{figure}
\hspace{-0.8cm}\epsfig{figure=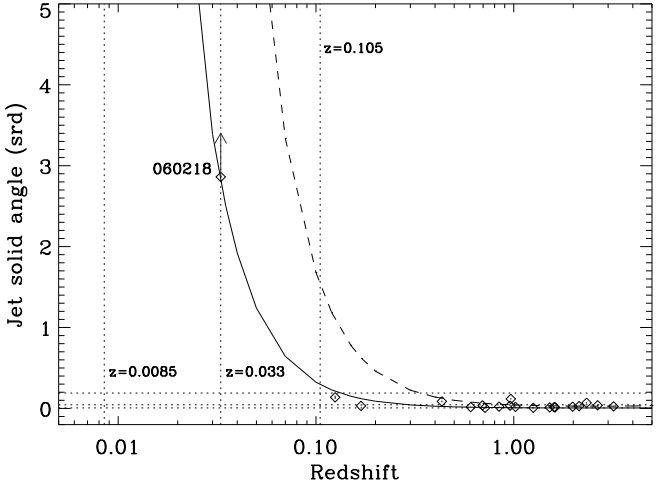,width=10cm}

\caption{GRBs jet solid angles as a function of redshift. Values are taken from
Nava et al. (2006), except for GRB\,060218 and GRB\,060614 (see text). 
The solid and dashed lines show, as a function of redshift, the jet solid angle required
to have a constant detection rate as a function of redshift by assuming
a homogeneous source distribution in time and space, a flat luminosity function
and neglecting detector's sensitivity threshold. The solid line is normalized to the
detection probability of a GRB with redshift and solid angle of GRB\,060218; the 
dashed line to the detection probability of a GRB lying at z = 1 and with jet
solid angle of 7$^{\circ}$.
The three vertical dotted lines correspond to the redshifts
of GRB\,980425, GRB\,060218 and GRB\,031203. The horizontal lines
delimitate the range of jet opening angles found for cosmological GRBs.
See text for details. 
}
\end{figure}

Several authors (e.g., \cite{Guetta04,Dellavalle06,Pian06,
Soderberg06b,Cobb06}) have pointed out
that sub--energetic GRBs may be the most frequent gamma-ray events in
the Universe. Indeed, since the volume sampled at $z$ = 0.033 is $10^4
\div 10^6$ times smaller than that probed by classical, distant GRBs,
the fact that we have observed one sub-energetic event out of $\sim$80
GRBs, with estimated redshift, indicates that the rate of these events
could be as large as $\sim$2000 GRBs Gpc$^{-3}$ yr$^{-1}$ (e.g.
\cite{Guetta06,Liang06,Soderberg06b}).  The hypothesis of
such high rate is further supported by the possibility, investigated
by Ghisellini et al (2006), that also GRB\,980425 and GRB\,031203 may
be truly sub--energetic events.\\ However in these estimates,
particular attention has to be paid to the collimation angle of the
emission. Indeed from radio afterglow modeling and no detection of the
jet break, it has been inferred that GRB060218 was much less
collimated than normal cosmological GRBs (see section 2). This
suggests that local sub-energetic GRBs are much less collimated than
the brighter and more distant ones. In this case, the larger jet solid
angle would, at least partly, compensate the smaller co-moving volume,
thus making the occurring rate of sub--energetic GRBs consistent with,
or not much higher than, that of bright cosmological GRBs (see 
\cite{Guetta06}).
However, by considering the combined effect of
the co--moving volume and jet opening angle on the detection
probability, it can be seen that the detection of local and quasi
spherical GRBs like GRB\,060218 is consistent with the hypothesis that
the jet angle distribution of local and distant GRBs is the same.
Indeed, by neglecting detector's limiting sensitivity and by assuming
a uniform jet, a homogeneous distribution in space, a rate independent
of redshift, and a flat luminosity function, the probability of
detecting a GRB lying at a redshift $z$ and emitting within a solid
angle $\Omega$ is
$$ \frac{dP(z,\Omega)}{dz} \propto \frac{\Omega}{4 \pi} \times 4 \pi
 \frac{dV_c}{dz} \propto \Omega \times \frac{dV_c}{dz} $$ where dVc is
 the co--moving volume element corresponding to the redshift interval
 ($z$ , $z + dz$) (e.g., \cite{Weinberg72,Peebles93}). The term
 $\Omega$ / 4$\pi$ accounts for the fact that the detection
 probability increases with increasing jet opening angle, and the term
 $4 \pi$ $dV_c/dz$ for the fact that for an uniform distribution the
 number of sources within $z$ and $z + dz$, and thus the detection
 probability (if neglecting detector's sensitivity limit), increases
 with redshift.  This is graphically shown in Fig.~3, where we plot
 the jet solid angles of GRBs in the sample of Nava et al. (2006) plus
 GRB\,060218 and GRB\,060614 (see Section 2), as a function of
 redshift. As can be seen, no trend in the jet angle distribution is
 apparent down to z$\sim$0.1--0.2, whereas there is a sudden increase
 in the jet opening angle for very low redshift if we include the
 lower limit to the collimation angle of GRB\,060218.  The solid and
 dashed lines show, as a function of redshift, the jet solid angle
 that a GRB must have in order to maintain constant P(z,$\Omega$).  The
 solid line is normalized to the redshift and jet angle lower limit of
 GRB\,060218, while the dashed line is normalized to the detection
 probability of a GRB with jet opening angle of 7$^{\circ}$ and
 located at a redshift of 1. Sources lying on the right line have a
 detection probability $\sim$5 times higher than those lying on the
 left one. For instance, at the redshift of GRB\,060218 a very weakly
 collimated emission is needed in order to have the same detection
 probability of a source with jet angle $\sim$10$^{\circ}$ located at
 $z$$\sim$0.2--0.3 (solid line), but even a spherical emission has
 $\sim$4 times lower detection probability of a source with a jet
 angle $\sim$5--10 degrees located at a redshift of $\sim$1--2 .  At
 the redshift of GRB\,980425 the detection probability is very low,
 compared to that of cosmological GRBs, even for spherical emission.
 Of course, things change if we include the possibility of detecting a
 source even when it is seen off--axis, which could be the case for
 GRB\,980425 and GRB\,031203 (as discussed above).  Thus, the very low
 redshift and (likely) wide jet opening angle of GRB\,060218, and also
 the possible off--axis detection of GRB\,980425, are consistent with
 the hypothesis that local GRBs have a jet angle distribution similar
 to that of distant GRBs.  A possible caveat with this scenario is the
 lack of detection of bright weakly collimated GRBs both in the local
 and high redshift universe. The most straightforward explanation is
 that there is a correlation between \eiso{} and jet opening angle, as
 it may be suggested by the narrow distribution of collimation
 corrected energies (\cite{Frail01,Berger03b, Ghirlanda04}). In this
 case, we would miss both close bright GRBs, because their narrow jet
 opening angle make their detection very unlikely (Fig.~3), and high
 redshift weakly collimated events, because they are the weaker ones
 and thus, due to the detectors sensitivity limits (not considered in
 Fig.~3), their detection probability quickly decreases with
 increasing redshift. \\ The main consequence of this scenario is that
 the occurrence rate of GRBs may be really as high as $\sim$2000
 GRBs Gpc$^{-3}$ yr$^{-1}$, both in the local Universe and at high
 redshift.

\subsection{The \epeiso{} plane and the GRB/SN connection}

From Fig.~1, one derives that all GRBs associated with 
SNe are consistent, or potentially consistent,
with the \epeiso{} correlation independently of their
\eiso{} or the SN peak magnitude and kinetic energy, with the
exceptions of GRB 980425 and possibly GRB\,031203. 
However, for these two events, and in particular for GRB\,980425, given its
very low redshift, the possibility that the deviation from the \epeiso{} correlation
is not real but due to an off--axis viewing angle cannot be excluded.  
Ghisellini et al. (2006) have proposed alternative explanations for the peculiar behavior in
the \epeiso{} plane of these two events.
One is the presence of scattering material of large
optical depth along the line of sight (which would have the effect of
decreasing the apparent \eiso{} and increasing the apparent \epi{}). As an
alternative, they suggest that, due to the limited energy band of the
instruments which detected them, the softest component of the prompt
emission of these two events was missed, leading to an overestimate of
\epi. The latter explanation is also supported by the fact that,
without the XRT (0.2--10 keV) measurement, the peak energy of
GRB\,060218 would have been overestimated and this burst would have been
classified as another outlier to the \epeiso{} correlation. 
Also, for GRB\,031203 there is possible evidence from the X-ray dust echo
measured by XMM (\cite{Vaughan04,Watson06}) that the soft prompt emission
component was missed by INTEGRAL/ISGRI, operating at energies above
$\sim 10$--15 keV. All these scenarios support the hypothesis that
the true \epi{} and \eiso{} of GRB\,980425 and GRB\,031203 are consistent with the 
\epeiso{} correlation, as all the other GRB/SN events.
However, one must caution
that both the inference that the \epi{} of GRB\,980425 could have been as low as
$<$1 keV (required to fit the \epeiso{} correlation) and the estimate of the prompt 
soft X--ray
flux of GRB\,031203 based on the dust echo are strongly model dependent.

The ``optical'' properties (i.e. luminosity at peak and expansion
velocities) of the SNe listed in Table~1 vary by a factor $\sim 5-10$ at
most, while the gamma-ray budget covers about 4 orders of
magnitude. This fact suggests that the difference between sub-energetic
and bright GRBs should not depend on the properties of SN
explosions, which are similar, but it is likely related to the
efficiency with which SNe are able to convert a significant fraction of
their kinetic energy into relativistic ejecta. 
However, after correcting
\eiso{} for the jet opening angle inferred from the break time in the
optical afterglow light curve, it is found that the GRB/SN events
GRB\,030329, GRB\,021211 and GRB\,050525A are characterized by radiated
energies, \ega, in the range ($\sim 0.5 \mbox{--} 1) \times 10^{50}$ erg
(see Table~1), showing that also for cosmological GRB/SN events the
energy radiated in gamma--rays may be only a small fraction of the SN
kinetic energy ($\sim 10^{51-52}$ erg). The location
of GRB/SN events in the $L_{\rm X,10}$ -- \epeiso{} plane (Fig.~2)
further supports the hypothesis that the emission mechanisms at play
are independent of the SN properties.

No significant correlations are found among the quantities reported
in Table~1. 
As can be seen in Fig.~4, 
no evidence of the correlation
between the peak magnitude of the SN and the \epi{} of the associated GRB 
reported by Li (2006) is apparent in our
enlarged sample. Even considering only the 7 GRBs associated with bright
SNe (filled circles), small variations of M$_{V}$ 
correspond to variations of \epi{} by a factor up to $\sim$100.
Thus, given also the scanty statistics, the existence of this correlation
cannot be currently supported.

\begin{figure}
\hspace{-0.8cm}\epsfig{figure=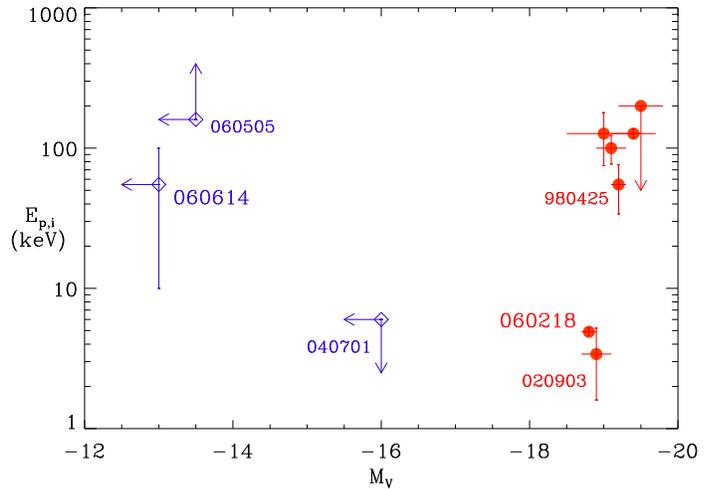,width=10cm}
\caption{Peak energy of GRB prompt emission vs. SN peak magnitude. 
Data are taken from Table~1. Diamonds represent GRBs with upper limits to
the luminosity of the possible associated SN. Filled circles are GRBs with
associated SNe. 
}
\end{figure}
\subsection{GRB\,060614: a different progenitor but similar emission mechanisms ?}

We find that the long lasting GRB\,060614, for which an association
with a bright Supernova can be excluded
(\cite{Dellavalle06b,Fynbo06,Gal-Yam06}), is also consistent with the
\epeiso{} correlation (see Fig. 1).  On the other hand, Gehrels
et al. (2006) have shown that, despite GRB\,060614 lasted more than
100s, it lies in the same region of the temporal lag -- peak
luminosity plane populated by short GRBs.  This evidence may point to
the existence of a class of GRBs with common properties and similar
progenitors, independently on their duration.  The fact that
GRB\,060614 follows the \epeiso{} correlation, while short events do
not, is a challenging evidence for this hypothesis (unless, as
discussed by Gehrels et al. 2006, one considers only the first pulse
of this event, which is characterized by values of \epi{} and \eiso{}
inconsistent with the correlation).\\ 
The inconsistency of short GRBs
with the \epeiso{} correlation may be explained, for instance, with
the relevant role of the circum--burst environment density and
distribution, which are expected to be very different in the merger
scenario (short GRBs) with respect to the collapse scenario (long
GRBs). The fact that both long GRBs associated with SN and long GRBs
without SN are consistent with the \epeiso{} correlation may suggest
that the circum--burst environment, the energy injection, or other
physical mechanisms at play are similar for the their
progenitors. This hypothesis is further supported by: {\sl i)} the location of
GRB\,060614 and XRF\,040701 in the \eiso{} -- $L_{\rm X,10}$ plane,
which is consistent (Fig.~2) with those of GRB/SN events (data for
XRF\,040701 and GRB\,060614 were derived from Fox et al. 2004 and
Mangano et al., paper in preparation, respectively); {\sl ii)} 
the fact that the existence of long lasting GRBs associated with very weak SNe may
still be explained with the explosion of massive progenitor stars (see
Della Valle et al. 2006; Tominaga et al. 2007 in preparation)
similarly to ``classical'' long-duration GRBs (e.g., \cite{Woosley06}).

\subsection{GRB\,060505}

Very low upper limits to the luminosity of an associated SN have also
been found for GRB\,060505 (\cite{Fynbo06}). Differently from
GRB\,060614 and XRF\,040701, this event is inconsistent with the
\epeiso{} correlation. One (unlikely) explanation for this behavior
could be that the association of this event with a galaxy at $z$=0.089
is not physical but due to chance superposition.  We computed the
track of GRB\,060505 in the \epeiso{} plane as a function of redshift
(see dotted curve in Fig.~2) and find that it is always outside the
$\pm$2$\sigma$ confidence region and that it would be marginally
consistent (i.e. within 99\% c.l.) with the \epeiso~ correlation for
$\sim2 < z < 6$ . It must be cautioned that the spectral information
provided by Swift/BAT for this event are rather poor and based on
survey mode data collected only up to 60 s after the GRB onset,
because Swift was approaching the South Atlantic Anomaly. In
addition, the short duration of this event, 4$\pm$1 s, combined with
its low fluence and hard spectrum (\cite{Hullinger06}) may indicate
that it belongs to the short GRB class, as also discussed by Fynbo et
al. (2006).  In this case the inconsistency with the \epeiso{}
correlation (which is not followed by short GRBs)
is not not surprising.

\section{Conclusions}

We analyzed and discussed the location in the \epeiso{} plane of two very 
interesting long GRBs: the local, sub--energetic GRB\,060218, associated with
SN2006aj, and GRB\,060614, for which an association with a bright
SN similar to other GRB-SNe can be 
excluded. We included in our analysis also other GRB/SN events and two
more GRBs with very deep limits to the magnitude of an associated SN.
The main implications of our analysis can be summarized as follows.

a) The consistency of GRB\,060218 with the \epeiso{} correlation favors
the hypothesis that this is a truly sub-energetic event rather than a
GRB seen off axis. The ratio between \eiso{} and $L_{\rm X,10}$ and the
radio afterglow properties of this event further support this
conclusion. If this is the case, GRB\,060218 can be considered as 
the prototype of a local sub--energetic GRB class. 

b) Based on simple
considerations on co--moving volume and jet solid angle effects on
GRB detection probability as a 
function of redshift, it is found that the detection of a close, weak
and poorly collimated (as suggested by modeling of radio data)
event like GRB\,060218 is consistent with the hypothesis that the rate and 
jet opening angle distributions
of local GRBs are similar
to those of cosmological GRBs. A correlation between jet opening angle and
luminosity can explain the lack of detection of local bright GRBs and
of distant, weakly collimated events. 
If this is the case, the occurrence rate of 
GRBs may be as high as $\sim$2000 GRBs Gpc$^{-3}$ yr$^{-1}$,
both in the local Universe and at high redshift.

c) All GRB/SN events are consistent with the \epeiso{} correlation, except for
GRB\,980425 and GRB\,031203. However, the first event is so close that
an off--axis detection is possible, whereas for the latter there are
observational indications that the \epi{} value could be consistent with the 
correlation. 
The consistency of GRB/SN events with the \epeiso{} correlation,
combined with energy budget considerations
and their location in the \eiso{} -- $L_{\rm X,10}$ diagram,
show that the emission properties of long GRBs do not depend on the 
properties of the associated SN. No clear evidence of correlation is found between
GRB and SN properties. in particular, all GRB/SN events seem to cluster in the
\epi{} - SN peak magnitude plane, with the only exception of GRB\,060218.

d) The consistency of GRB\,060614 with the \epeiso{} correlation
shows that the emission mechanisms at play in long GRBs may be
independent from the progenitor type. GRB\,060505, another GRB with
stringent upper limits to the luminosity of an associated SN, is
inconsistent with the \epeiso{} correlation. However, the short
duration, low fluence and hard spectrum of this event may suggest that
it belongs to the short GRBs class.


\end{document}